# Design of Mathematical Model for Prediction of Mechanical Properties of Bar at Strain-Heat Hardening

**Azamat A. Kanayev**

At research of influencing for an elemental composition on mechanical properties of the hot-rolled and strain-heat hardened rebar with dia of 14 mms was taken to base himself on low carbon steel with carbon content from 0,15% to 0,21%. (tab. 1). Such level of carbon content allows to receive rather broad temperature range of a self-tempering without essential weakening of steel. It is of great importance for providing with stability of technological process for straining-heat hardening on the assumption of fluctuation in temperatures for a self-tempering lengthwise of 70-75-m bars. It is not less the important role the indicated interval of carbon content has for structure making at straining-heat treatment the high-strength and enough the plastic self-tempered martensite with avoidance of danger of cracking and accordingly by advancement of resistance to brittle fracture. The content of other constant admixtures (Mn, Si, S, P) in investigated steels is changed in quite broad ranges: Mn from 0,54% to 2,05%; Si from 0,03% to 1,49%; S from 0,014% to 0,095%; P from 0,08% to 0,023%. It is known, that in connection with relative accessibility and cheapness the most broad using as alloying element have received manganese and silicon.

Table 1 Mechanical Properties of Hot-Rolled Reinforcing Bar Steels and Elemental Composition for Investigated Steels, %

| Melt | C | Mn | Si | S | P | Mechanical properties | | |
|---|---|---|---|---|---|---|---|---|
| | | | | | | $\sigma_y$,MPa | $\sigma_s$,MPa | $\delta_5$,% |
| 1 | 0,19 | 0,54 | 0,22 | 0,030 | 0,016 | 293 | 443 | 40,20 |
| 2 | 0,15 | 0,73 | 0,24 | 0,029 | 0,023 | 303 | 466 | 40,10 |
| 3 | 0,21 | 1,18 | 0,24 | 0,035 | 0,018 | 348 | 550 | 36,10 |
| 4 | 0,16 | 2,05 | 0,24 | 0,035 | 0,020 | 368 | 604 | 29,90 |
| 5 | 0,19 | 0,95 | 0,66 | 0,024 | 0,008 | 359 | 545 | 34,50 |
| 6 | 0,19 | 1,10 | 0,96 | 0,014 | 0,020 | 366 | 539 | 35,70 |
| 7 | 0,19 | 1,40 | 1,08 | 0,030 | 0,019 | 401 | 593 | 33,70 |
| 8 | 0,19 | 1,24 | 1,49 | 0,021 | 0,017 | 419 | 613 | 32,70 |
| 9 | 0,20 | 1,84 | 0,65 | 0,044 | 0,022 | 431 | 635 | 32,50 |
| 10 | 0,20 | 1,13 | 0,03 | 0,095 | 0,014 | 412 | 611 | 29,00 |

Valuable advantage of manganese is its ability actively to increase a hardenability of steel, that is explained by a deceleration of decay of undercooled austenite both in perlitic, and in intermediate temperature range. Nevertheless, it is not common opinion on influencing manganese on structure and the properties of steel for a different elemental composition. There are values as about positive, and neutral or even negative influencing of manganese, as alloying element, on properties of steel [1, 2]. From values of tab. 1 follows, that in a hot-rolled state the increase of a content of manganese from 0,54% to 2,05% at a practically identical content of silicon (0,22%-0,24%) raises a yield point and breaking point accordingly from 293 MPa and 443 MPa to 368 MPa and 604 MPa. The aspect ratio for at that is reduced from 40,2% to 29,9%. It is answered with data of research works [3], in which one is fixed, that the raise of a content of manganese to 2% hardens ferrite and decreases turning for generation of microfractures in colonys of a perlite in hot-rolled low carbon steel. The tendency is viewed, that the increase of a content of manganese (at rather a high level of silicon 0,65%-0,66%) essentially hardenes a steel. So, the increase of a content of manganese from 0,95% (at 0,66% Si) to 1,84% (at 0,65 % Si) causes the growth of a yield point and breaking point accordingly from 359 MPa and 545 MPa to 431 MPa and 635 MPa. The aspect ratio $d_5$ changes for all that unsignificantly, reduction from 34,5 % to 32,5%. As against these data Kugushin and another researchers have determined, that the raise of a content of manganese to 2,0 % does not influence neither mechanical properties, nor on critical transition temperature to a brittle state of an air-hardened steel [4]. In article [5] the presence Ìn at steels is negatively evaluated and is pointed to the considerable drop of a plasticity with growth of a content of manganese in steel from 0,35 % C.
Silicon, as well as the manganese, intricately and ambiguously influences on plasticity and strength of steel. The influencing is changed depending on a content of the silicon and other chemical elements in steel. Silicon, apart from ability actively to deoxidate steel at the expense of transmission to valence electrons from an last shell $3s^2p^2$ to atoms of oxygen, having the last electronic shell $2s^2p^4$ with reaching by them as a result of stable electronic configurations $2s^2p^6$. And also, silicon increases stability of a martensite against tempering. The most of researchers state common opinion, that silicon at content to 1,5 %, as well as the manganese, influences hardening action on steel. For all that plasticity of steel practically is not gone to worth. These clauses are greatly correspond to our value, viewed in tab. 3.1. The growth of a content of silicon from 0,22 % to 1,49 % carries on to increase of breaking point with 443 MPa to 613 MPa, and the aspect ratio for at that is reduced unsignificantly from 40,2 % to 32,7%. The silicon, specially in a complex with manganese, ensures the considerable hardening at saving a plasticity (tab. 3.1). Such influencing of silicon is valid at carbon content in steel above 0,21-0,25%. A similar action of silicon defined in research work [6]. There are informations, that the silicon (to 2,0 %) drops quadrangleness of a lattice of an initial martensite and reduces turning for formation of quenching cracks, as decreases a strain of a test piece at quenching [7]. At the same time it is necessary to mark and other estimations of influencing of silicon on properties of steel. The researchers [8] mention a raise of critical temperature of transition to a brittle state of the normalized building steel with 1,46 % Si. The negative influencing of silicon above 0,5 % on end rolling temperature of hot-rolled building steel is pointed to as well. N. Livschits, A. Rakhmanov, N. Sitnickov [9] arrived to a conclusion, that already at a content above 0,37 % silicon influences harmfully on spreading of cracks in the normalized and improved steels with 0,15-0,20 % C. The analysis of results given above in research works and different authors is been evidence of the influencing of manganese and silicon on mechanical properties for low carbon  steel is ambiguous. It is in complex dependence on carbon content and other chemical elements in steel, its process engineering effecting and heat treatment.

For definition of dependence between mechanical properties, on the one hand, and elemental composition, with other hand, have used a method of the multiple regression analysis. It allows to select the most significant magnitudes describing a yield point, breaking point and aspect ratio, to estimate a degree of influencing on these properties of separate chemical elements and technological parameters. The multiple regression analysis is successfully applied to study of communications between one dependent and several independent variables [10, 11, 12]. A common computational problem, which one solve at the analysis by a method of a multiple regression, consists in substitution of a straight line (or plane  in m-dimensional space, where m - number of independent variables) to some range of points. Taking advantage vectorial table of symbols, we shall give to Y - a vector of the case consisting from n of elements, through x - matrix of independent variables by a size m x n, where m - number of independent variables, and n - number of cases. In these table of symbols the problem may be formulated as follows:

$$Y = X\beta + \varepsilon, \qquad (1)$$

Here $\varepsilon$ there are independent random errors with average $\theta$, which one are interpreted as an error of cases, and $\beta$ - vector of unknown parameters, which it is necessary to estimate. Estimation of parameters $\beta$ we shall give to b. In the given problem an dependent variable - mechanical properties ($\sigma_b$, $\sigma_y$, $\delta_5$), and independent variables - content of constant admixtures in steel (C, Mn, Si, S, P). The dependence between variables is supposed linear. Dependent variable is named as response, and independent variables - predictors or control variables. This terminology emphasizes, that a series of variables influences on one variable - response. Let's remark also, that the equation (1) allows to find an estimation of response at any value of control variables.

Before immediate application one or another statistical analysis often there is necessary to transformate magnitude of input values. So, for doing regression of the analysis it is necessary to take the logarithm for a yield point, breaking point and aspect ratio. It stabilizes a variance and often is applied in solution of similar kind of problems. It can be interpreted as follows: the higher is the absolute value of variable, the above level of random errors. At a taking the logarithm all errors become approximately identical. Therefore find a linear dependence not between percentage of the appropriate chemical elements in investigated steel, on the one hand, and strength, plastic characteristics, on other hand, but dependence of percentage contents of elements and logarithmical characteristics of mechanical properties of rebars, receiving more stable estimations of parameters of model. Afterwards, when the model will be constructed, it is possible to proceed to basic values. The problem consists in design of model:

$$Ln(\sigma_y) = b_0 + B_1*C + b_2*Mn + b_3*Si + b_4*S + b_5*P, \quad (2)$$

Here $b_1$ - unknown coefficient; $b_0$ - absolute term (also is unknown).
For all that, the model for yield point:

$$\sigma_y = \exp\{b_0 + B_1*C + b_2*Mn + b_3*Si + b_4*S + b_5*P\}, \quad (3)$$

These equations allow to estimate unknowns parameters and to test significance of regression and adequacy to designed model to input values. The procedure both for breaking point and aspect ratio is similar.
After transformation the tab. 1 viewed as follows (tab. 2):

Table 2 Input Data for Regression Analysis (Hot-Rolled State)

| Melt | Mechanical Properties | | | | | |
|---|---|---|---|---|---|---|
| | before taking the logarithm | | | after taking the logarithm | | |
| | $\sigma_y$, MPa | $\sigma_b$, MPa | $\delta_5$,% | Ln ($\sigma_y$) | Ln ($\sigma_b$) | Ln ($\delta_5$) |
| 1 | 293 | 443 | 40,20 | 5,680 | 6,094 | 3,694 |
| 2 | 303 | 466 | 40,10 | 5,714 | 6,144 | 3,691 |
| 3 | 348 | 550 | 36,10 | 5,852 | 6,310 | 3,586 |
| 4 | 368 | 604 | 29,90 | 5,908 | 6,404 | 3,398 |
| 5 | 359 | 545 | 34,50 | 5,883 | 6,301 | 3,541 |
| 6 | 366 | 539 | 35,70 | 5,903 | 6,290 | 3,575 |
| 7 | 401 | 593 | 33,70 | 5,994 | 6,385 | 3,517 |
| 8 | 419 | 613 | 32,70 | 6,038 | 6,418 | 3,487 |
| 9 | 431 | 635 | 32,50 | 6,066 | 6,454 | 3,481 |
| 10 | 412 | 611 | 29,00 | 6,021 | 6,415 | 3,367 |

The problem of the given investigation consists in design of linear regression between dependent variables $\sigma_b$, $\sigma_y$, $\delta_5$ = Ln ($\sigma_b$), Ln ($\sigma_y$), Ln ($\delta_5$) and independent variables C, Mn, Si, S, P.
In fig. 1 the normal probability plot for carbon is shown normally, which one is applied for an estimation of a normality distribution for variable, i.e. closeness of this distribution to normal. The dependence between a selected variable and expected values, received from normal distribution, is given in scatterplot.
At first for design of standard normal probability plot all values are arranged on rank. On these ranks is calculated the values z (i.e. standardized values of normal distribution) in the supposition, that the values have normal distribution. These values z are put aside on Y-axis of the plot. If the observed values (put aside on X-axis) are distributed normally, all values on the plot put to straight line. If the values are not normally distributed, they will be changed form of line. On this plot it is possible easily to find deviations. If the obvious noncoincidence is viewed, and the values place concerning a line definitely (for example, as S letter), then it is possible to apply other transformation. For evaluation on ranks of expected normal probability values, i.e. corresponding to normal z-values, the following formula will be used:

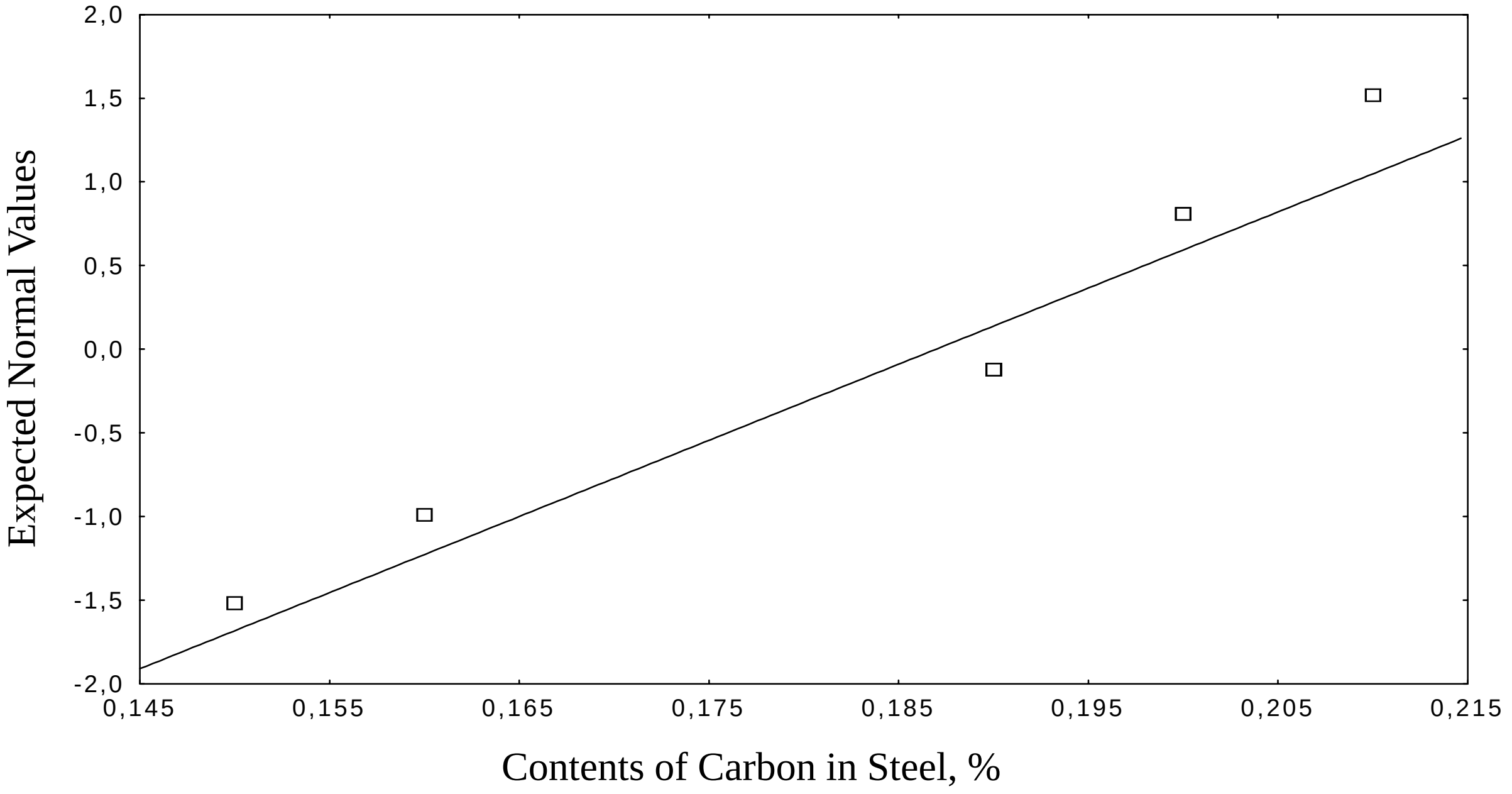

Fig. 1 Normal Probability Plot for Carbon, y=-8.682+47.165*x+eps

$$Z_j = F^{-1} [(3*j-1) / (3*N + 1)], \quad (4)$$

Here $F^{-1}$ there is an inverse function of normal distribution (converting normal probability p in a normal value z)

$Z_j$ - normal probability value for j-al value (rank) of variable with N cases.

As displays the given plot on fig. 1, the carbon content in rebar satisfactorily corresponds to normal distribution.

On 2D Box plots is viewed the rectangles and lines. This type of the block statistical plots represents the box plot showed in fig. 2 for medians (and also minimum and maximum values and 25-th and 75-th percentiles) for columns or lines of the block.

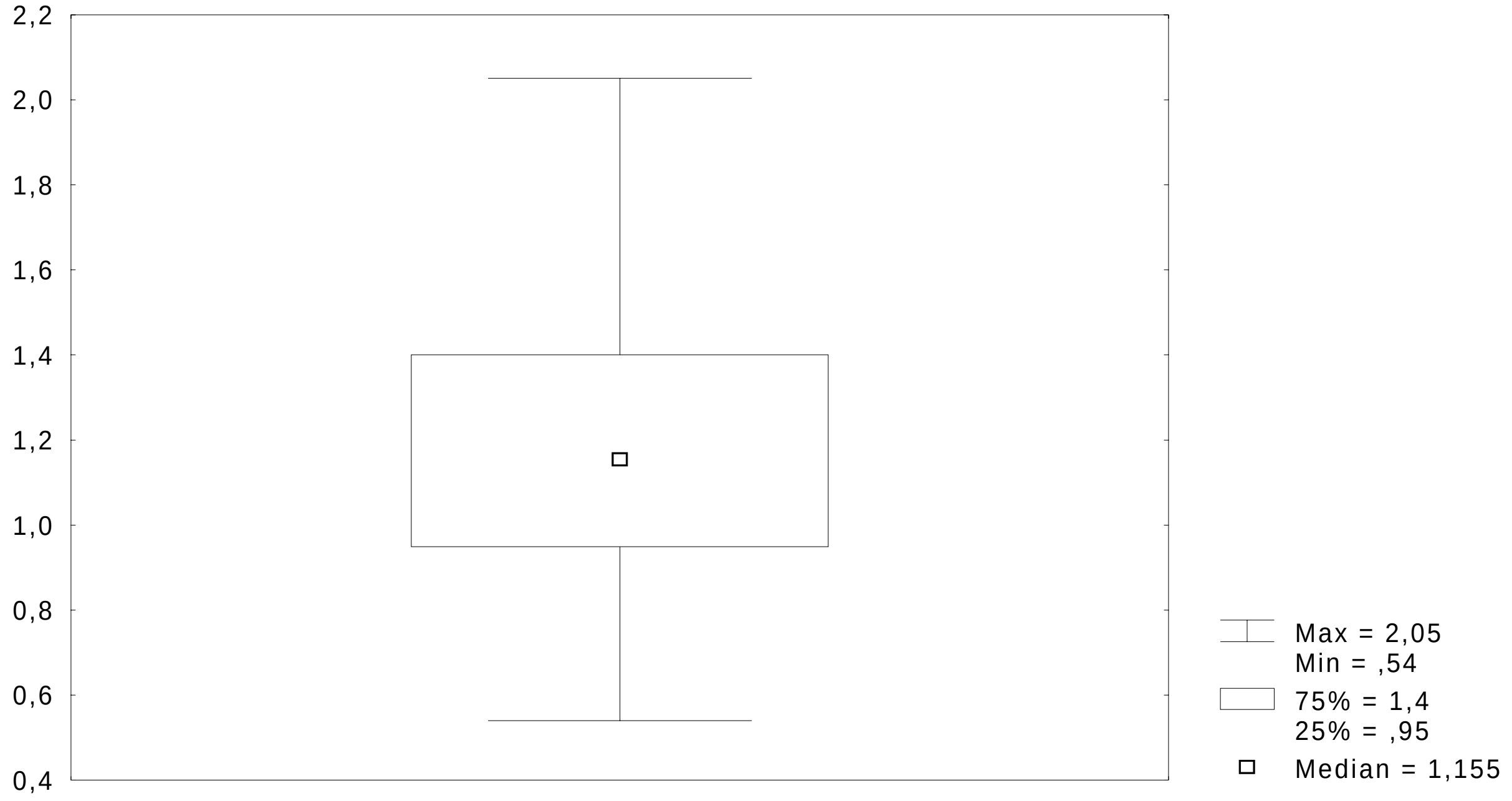

Fig. 2 2D Box Plot (Range of Manganese Values) - Rectangles - Lines.

Each rectangle maps values from one column either line or end line, which one are arranged outside of rectangle and pointing to selected range as well.
2D Histogram (fig. 3) are plot representations of distribution of frequencies for selected variables. For them is drawn the column for each interval (class), its height is proportional to frequency of class.
At transition from plots to regression analysis we shall take one of the methods of step by step regression. It consists as follows: for each step in model is included or excluded some independent variable. Thus, it is selected the great number of “significant” variables. It allows to reduce number of variables, which one describe the dependence. In this case we select step by step method of insert. At using of this method in regression equation sequentially are included independent variables, while the equation will not become satisfactorily to describe input values. The insert of variables is determined with aid of F-test. After STATISTICA will do calculations, the output window of the analysis will appear on the screen, which one has following simple structure: the upper part of window - information and lower part allows with all side to show the results of the analysis (fig. 5).
On fig. 5 Std. Error of Estimate - this statistic is standard measure of scattering for observed values from regression straight line;
- Intersept is value of coefficient $b_0$ in the regression equation;
- Std. Error is standard error of coefficient $b_0$ in the regression equation;
T-test will be used for check of hypothesis about equality to 0 absolute terms of regression. p is significance level.
As is well known, the statistical significance of result is an estimated measure of confidence in its “truth” (in sense “representative sampling”). The p-level is an index were in descending dependence on reliability of result. The higher p level the lower level of confidence to the dependence between variables, finding out in sampling. Just, the p-level represents probability of an error, concerned with distribution of observed result to whole sampling. Usually in many areas the result p = 0,05 is reasonable limit of statistical significance. However it is necessary to remember, that this level still includes quite high probability of an error (5 %). The results, where significant level p = 0,01 usually are considered as statistically significant, and results with p level equals 0,005 or p = 0,001 as high significant. We have coefficient of determination ($R^2$) in the equation of regression equals 0,973. It means, the designed regression explains 97,3 % of scatter of values Ln ($\sigma_y$) from average. It is good result. While a high value F-test = 45,0182 (test used for check of significance of regression) and the significance level p=0,0004, show the designed regression is highly significant with probability of an error 0,04%.

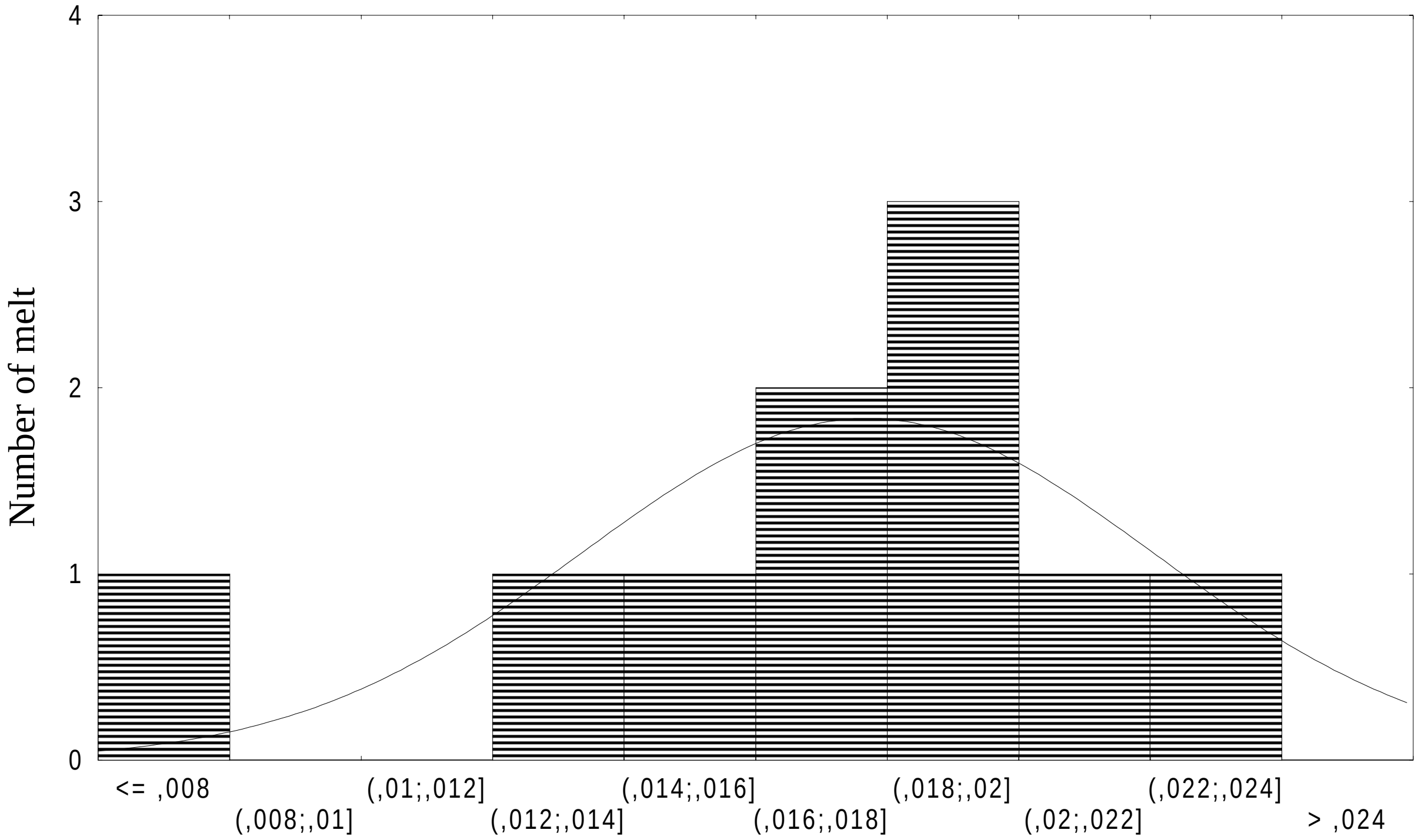

Fig.4 Histogram for phosphorus {y=10 * 0.002 * normal (x, 0.0177; 0.0043474)}

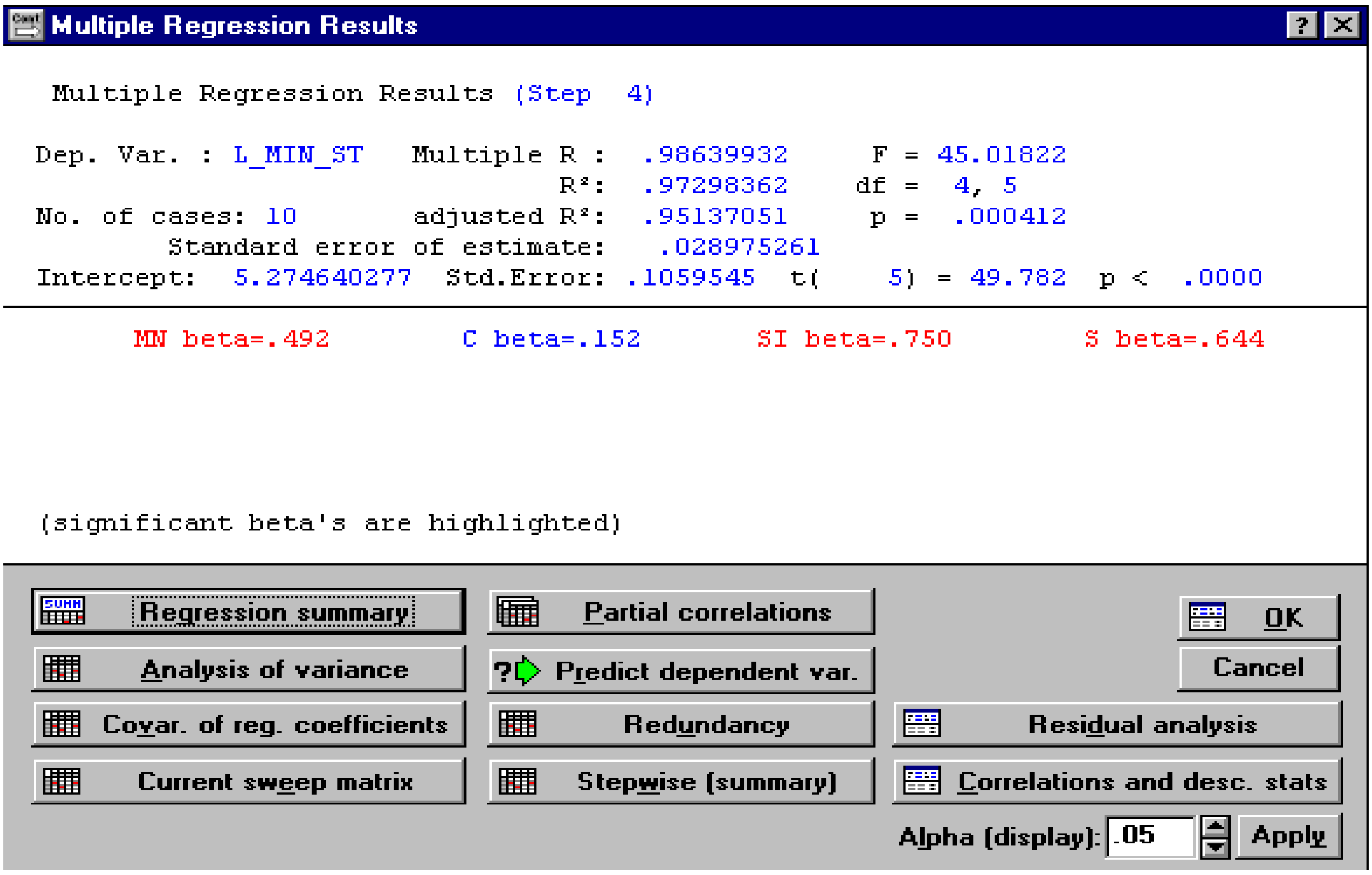

Fig. 5 Window with Results of Analysis.

In tab. 3 the results of the analysis for yield point are given, on which one it is possible to write the equation of regression.

In the fourth column (tab. 3) the required coefficients are arranged. The equation of regression it follows:

$$\mathrm{Ln}\,(\sigma_y) = 5{,}5414 + 0{,}2553*Si + 6{,}4254*S + 0{,}1145*Mn - 7{,}3137*P \qquad (5)$$

After transition to observed variables model we have:

$$\sigma_y = \mathrm{Exp}\,(5{,}5414 + 0{,}2553*Si + 6{,}4254*S + 0{,}1145*Mn - 7{,}3137*P) \qquad (6)$$

Table. 3 Results of Regression for Yield Point (Hot-Rolled State)

| Quantity of melt, #10 | BETA | St. Error of BETA | Regression Constant b | St. Error of b | t - Student Distribution | p level |
|---|---|---|---|---|---|---|
| Intercept | | | 5.54143 | .112092 | 49.43648 | .000000 |
| SI | .848726 | .175549 | .25526 | .052798 | 4.83469 | .004737 |
| S | .401125 | .193593 | 6.42536 | 3.101046 | 2.07200 | .093003 |
| MN | .426541 | .185473 | .11453 | .049803 | 2.29975 | .069794 |
| P | -.288849 | .158669 | -7.31367 | 4.017519 | -1.82045 | .128336 |

The qualitatively designed equation may be interpreted as follows: the yield point grows with increase of content of silicon and manganese, and also is reduced with growth of phosphorous content. It is possible to explain some exceptions, for instance, influencing of content of sulfur on $\sigma_y$ by correlating influence of other factors.
Let's recount values Ln ($\sigma_y$) on the assumption of designed model for different values of independent variables C, Mn, Si, S, P (tab 4).

Table 4 Results of the Multiple Regression Analysis (Hot-Rolled State).

| Quantity of Melt | Observed Value | Predicted Value | Residual | Standard Predicted Value | Standard Residual | Standard Error of Predicted Value | Mahalanobi-se Distance | Deleted Residual | Cook's Distance |
|---|---|---|---|---|---|---|---|---|---|
| 1 | 5,680172 | 5,716630 | -0,036457 | -1,45834 | -1,18800 | 0,020190 | 2,995795 | -0,064283 | 0,316565 |
| 2 | 5,713733 | 5,693855 | 0,019877 | -1,63381 | 0,64773 | 0,027625 | 6,393281 | 0,104819 | 1,575720 |
| 3 | 5,852202 | 5,848717 | 0,003485 | -0,44066 | 0,11357 | 0,023423 | 4,343255 | 0,008349 | 0,007188 |
| 4 | 5,908083 | 5,925167 | -0,017084 | 0,14836 | -0,55671 | 0,028347 | 6,779560 | -0,116444 | 2,047591 |
| 5 | 5,883322 | 5,860600 | 0,022723 | -0,34911 | 0,74045 | 0,027526 | 6,341111 | 0,116269 | 1,924901 |
| 6 | 5,902633 | 5,885355 | 0,017278 | -0,15837 | 0,56302 | 0,016253 | 1,624562 | 0,024014 | 0,028627 |
| 7 | 5,993961 | 6,015608 | -0,021647 | 0,84518 | -0,70539 | 0,015514 | 1,400092 | -0,029078 | 0,038244 |
| 8 | 6,037871 | 6,047334 | -0,009463 | 1,08961 | -0,30837 | 0,023220 | 4,252536 | -0,022137 | 0,049650 |
| 9 | 6,066108 | 6,047124 | 0,018984 | 1,08799 | 0,61863 | 0,020903 | 3,275688 | 0,035416 | 0,102993 |
| 10 | 6,021023 | 6,018720 | 0,002303 | 0,86915 | 0,07505 | 0,029813 | 7,594119 | 0,040974 | 0,280424 |
| Minimum | 5,680172 | 5,693855 | -0,036457 | -1,63381 | -1,18800 | 0,015514 | 1,400092 | -0,116444 | 0,007188 |
| Maximum | 6,066108 | 6,047334 | 0,022723 | 1,08961 | 0,74045 | 0,029813 | 7,594119 | 0,116269 | 2,047591 |
| Average | 5,905911 | 5,905911 | -0,000000 | -0,00000 | -0,00000 | 0,023281 | 4,500000 | 0,009790 | 0,637190 |
| Median | 5,905358 | 5,905261 | 0,002894 | -0,00501 | 0,09431 | 0,023321 | 4,297895 | 0,016182 | 0,191708 |

In tab. 4 the residual are differences between observed values and values, forecasted by investigated model. The better model is corresponded with values, the less magnitude of residuals. I-al residual ($e_i$) is evaluated as:

$$E_i = (y_i - y_p) \qquad (7)$$

where $y_i$ - observed value;
$y_p$ - respective predicted value.
The standardized values of residual are computed as a difference between the observed and predicted values, divided by square root from root-mean-square value of residual. The independent variables in the regression equation may be showed by points in multidimensional space (each case is pictured by point). In this space it is possible to put point of centre. This “midpoint” in multidimensional space is called as centroid, i.e. centre of gravity. Mahalanobise distance is determined as distance from an observed point to center of gravity in multidimensional space, defined correlated (unorthogonal) independent variables (if the independent variables are uncorrelated, Mahalanobise distance being congruents with usual Euclidean distance). This measure allows, in particular to define whether the given case a deviation with respect to rest values of independent variables.

The deleted residual are values of residual for the corresponding cases, which one were excluded from process of regression analysis. If the deleted residual considerably differs from the corresponding standardized value of residual, probably, this case is a deviation, because its exception essentially changes the regression equation. One more measure of influencing of the corresponding case for the regression equation is Cook's distance. Its value shows a diffrences between calculated B - coefficients and values, which one would be received by exception of the corresponding case. In adequate model all Cook's distances should be approximately identical; if it is not, it may be consider the relevant case (or cases) displaces the estimation of regression coefficients.

Investigating residual, it is possible to estimate a degree of adequacy to model. STATISTICA allows to view residual of model both in graphic presentation, and in spreadsheets. For estimation of adequacy to model it is best to use visual methods, for all that we shall consider normally probability plot of residual (fig. 6).

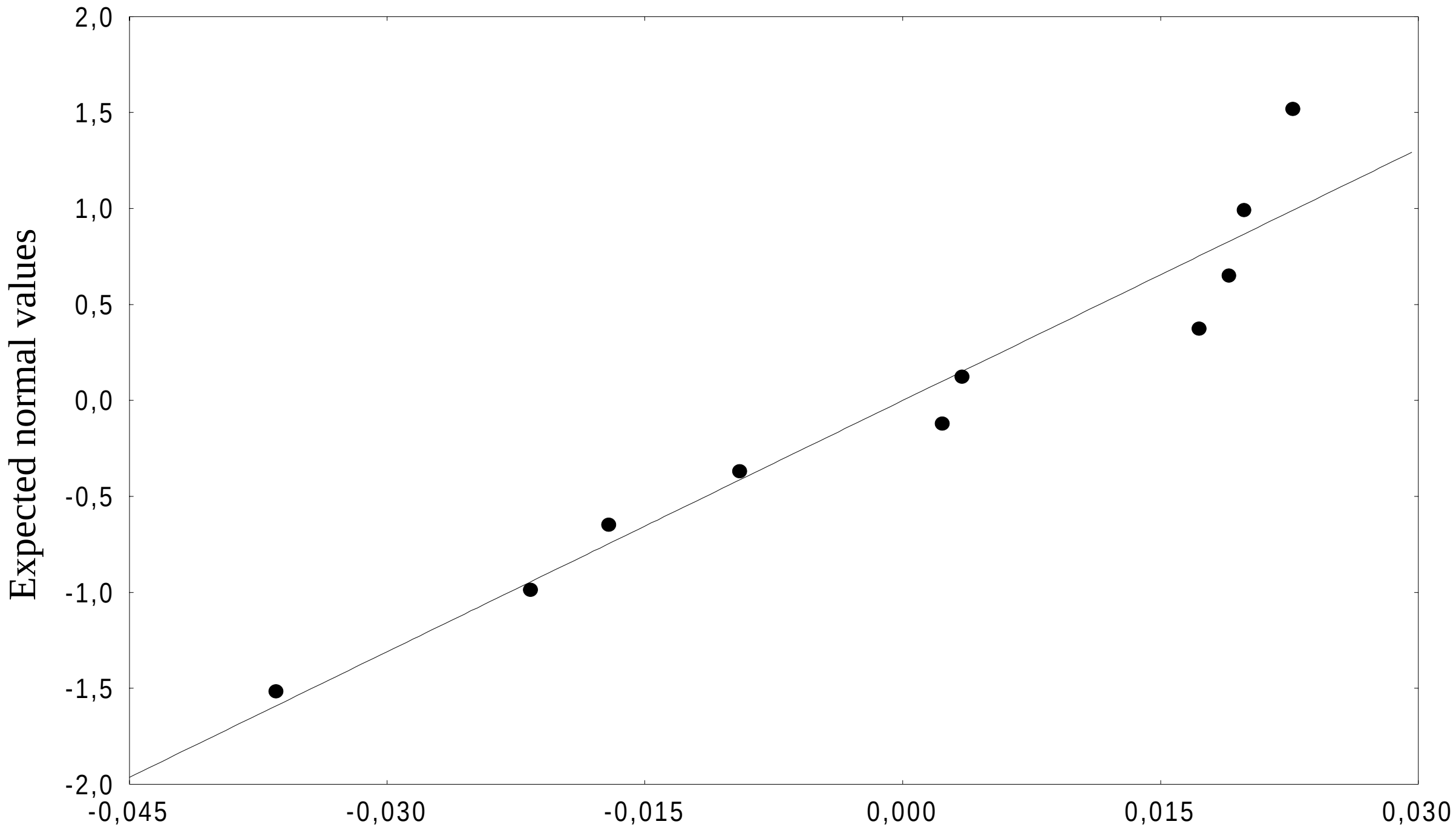

Fig. 6 Normal Probability Plot of Residual

The plot of residual shows, it is rather good puts on line, which one meet to the normal distribution. Therefore supposition about normal distribution of errors is carried out.

Other types of the residual plots for the given model are given below as well. (fig. 7-8).

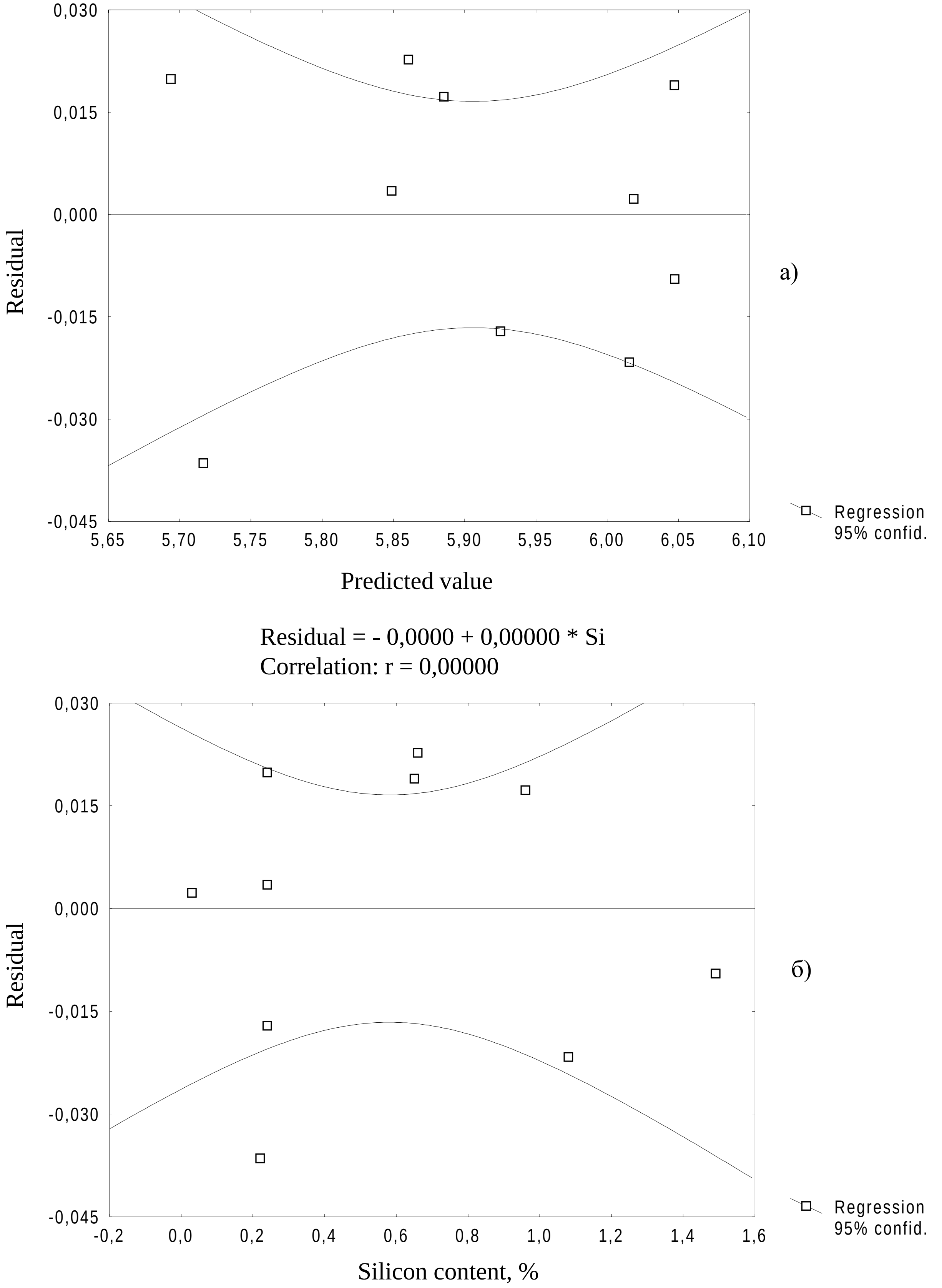

Fig. 27 Plots of residual in linear model of dependence for log of yield point a) on the predicted values; b) on variable (Si).

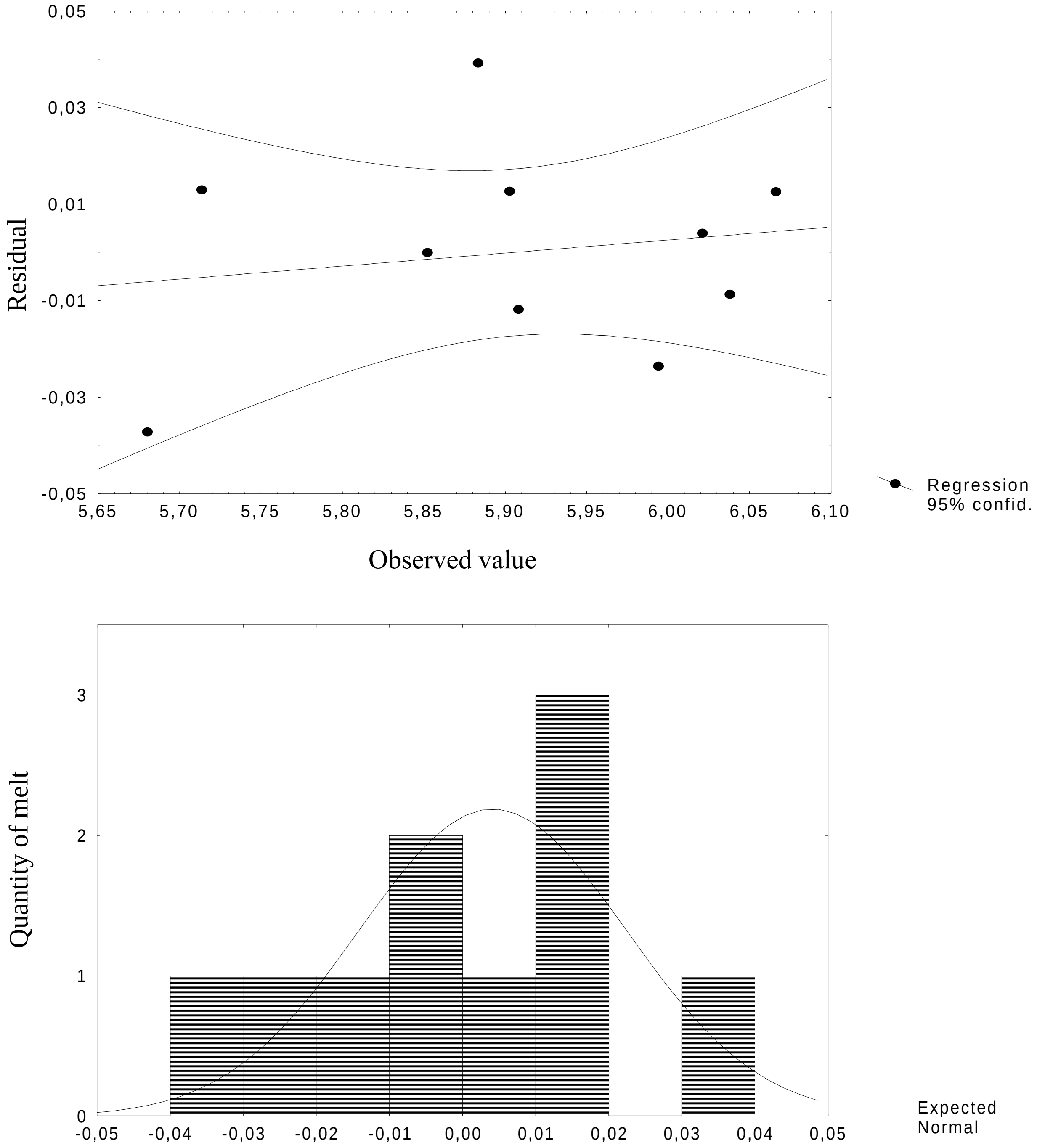

Fig. 8 Plots of residual in linear model of dependence for log of yield point a) on observed values; b) on residual distribution.

According to the described process are done statistical analysis both data processing for breaking point and aspect ratio of investigated steels for strain - heat hardened state. For all that took into consideration the discharge and pressure of cooling water, rolling rate ($\upsilon$), temperature of the end of strain, pause time and water temperature (tab. 5-8).

Applying of the multiple regression analysis to quantitatively estimate the complex of factors, influencing on $\sigma_y$, $\sigma_b$ and $\delta_5$ for strain-heat hardened rebar from 35GS (grade of steel) with dia 14 mms have allowed to receive the following equations of regression:

Table 5 Technological Factors of Strain-Heat Hardened Rebars and Mechnical Properties for Investigated Steels

| Number of Melt | Technological Factors | | | | | Mechanical properties | | |
|---|---|---|---|---|---|---|---|---|
| | $Q, m^3/h$ | $P$, MPa | $t_{e.r.}$, °C | $\Delta\tau$, s | $\upsilon$, m/s | $\sigma_y$, MPa | $\sigma_b$, MPa | $\delta_5$, % |
| 1 | 349 | 0,40 | 851 | 0,20 | 7,60 | 303 | 466 | 29,7 |
| 2 | 411 | 0,55 | 897 | 0,57 | 8,20 | 351 | 512 | 23,1 |
| 3 | 398 | 0,72 | 885 | 0,63 | 7,90 | 447 | 605 | 16,3 |
| 4 | 453 | 0,77 | 932 | 0,58 | 7,95 | 458 | 619 | 14,1 |
| 5 | 396 | 0,81 | 995 | 0,62 | 8,13 | 451 | 611 | 15,7 |
| 6 | 408 | 0,83 | 967 | 1,00 | 8,20 | 601 | 787 | 11,0 |
| 7 | 397 | 0,79 | 959 | 3,10 | 7,80 | 607 | 762 | 12,4 |
| 8 | 391 | 0,77 | 963 | 2,10 | 9,70 | 585 | 676 | 12,0 |
| 9 | 402 | 0,82 | 966 | 1,95 | 14,10 | 479 | 671 | 10,7 |
| 10 | 448 | 0,80 | 961 | 1,97 | 10,30 | 685 | 833 | 8,5 |
| 11 | 555 | 0,78 | 964 | 2,20 | 10,70 | 796 | 989 | 8,2 |
| 12 | 603 | 0,77 | 968 | 1,95 | 14,20 | 784 | 977 | 9,3 |

Table 6 Results of Regression for Yield Point (Heat Hardened State)

| Quantity of Melt, #10 | BETA | St. Error of BETA | Regression Constant b | St. Error of b | t - Student Distribution | p level |
|---|---|---|---|---|---|---|
| Intercept | | | -0,009228 | 1,419001 | -,006503 | 0,994971 |
| $\delta_5$ | 0,399234 | 0,138151 | 0,131995 | 0,045676 | 2,889836 | 0,020204 |
| Ln(Q) | 0,467955 | 0,125188 | 0,907433 | 0,242757 | 3,738031 | 0,005721 |
| P | 0,339816 | 0,138957 | 0,796075 | 0,325530 | 2,445475 | 0,040220 |

## Table 7 Initial Data for Resression Analysis (Heat Hardened State)

| Number of Melt | Mechanical Properties | | | | | | Technological Factors | | | | | |
|---|---|---|---|---|---|---|---|---|---|---|---|---|
| | before taking the logarithm | | | after taking the logarithm | | | before taking the logarithm | | | after taking the logarithm | | |
| | $\sigma_y$,MPA | $\sigma_b$, MPa | $\delta_5$, % | Ln ($\sigma_y$) | Ln ($\sigma_b$) | Ln ($\delta_5$) | Q, $m^3/h$ | $t_{e.r.}$, °C | $\upsilon$, m/s | Q, $m^3/h$ | $t_{e.r.}$, °C | $\upsilon$, m/s |
| 1 | 303 | 466 | 29,7 | 5,714 | 6,144 | 3,391 | 349 | 851 | 7,60 | 5,855 | 6,746 | 2,028 |
| 2 | 351 | 512 | 23,1 | 5,861 | 6,238 | 3,140 | 411 | 897 | 8,20 | 6,019 | 6,799 | 2,104 |
| 3 | 447 | 605 | 16,3 | 6,103 | 6,405 | 2,791 | 398 | 885 | 7,90 | 5,986 | 6,786 | 2,067 |
| 4 | 458 | 619 | 14,1 | 6,127 | 6,428 | 2,646 | 453 | 932 | 7,95 | 6,116 | 6,837 | 2,073 |
| 5 | 451 | 611 | 15,7 | 6,111 | 6,415 | 2,754 | 396 | 995 | 8,13 | 5,981 | 6,903 | 2,096 |
| 6 | 601 | 787 | 11,0 | 6,399 | 6,668 | 2,398 | 408 | 967 | 8,20 | 6,011 | 6,874 | 2,104 |
| 7 | 607 | 762 | 12,4 | 6,409 | 6,636 | 2,518 | 397 | 959 | 7,80 | 5,984 | 6,866 | 2,054 |
| 8 | 585 | 676 | 12,0 | 6,372 | 6,516 | 2,485 | 391 | 963 | 9,70 | 5,969 | 6,870 | 2,272 |
| 9 | 479 | 671 | 10,7 | 6,172 | 6,509 | 2,370 | 402 | 966 | 14,10 | 5,996 | 6,873 | 2,646 |
| 10 | 685 | 833 | 8,5 | 6,529 | 6,725 | 2,140 | 448 | 961 | 10,30 | 6,105 | 6,868 | 2,332 |
| 11 | 796 | 989 | 8,2 | 6,680 | 6,897 | 2,104 | 555 | 964 | 10,70 | 6,319 | 6,871 | 2,370 |
| 12 | 784 | 977 | 9,3 | 6,664 | 6,884 | 2,230 | 603 | 968 | 14,20 | 6,402 | 6,875 | 2,653 |

## Table 8 Results of Multiple Regression Module (Heat Hardened State)

| Quantity of Melt | Observed Value | Predicted Value | Residual | Standard Predicted Value | Standard Residual | Standard Error of Predicted Value | Mahalanobise Distance | Deleted Residual | Cook's Distance |
|---|---|---|---|---|---|---|---|---|---|
| 1 | 5,713733 | 5,648688 | 0,065045 | -2,15075 | 0,57408 | 0,095807 | 6,948477 | 0,228238 | 0,725357 |
| 2 | 5,860786 | 5,965322 | -0,104536 | -1,03967 | -0,92262 | 0,060183 | 2,186856 | -0,145621 | 0,116512 |
| 3 | 6,102559 | 6,079408 | 0,023150 | -0,63934 | 0,20432 | 0,047173 | 0,990099 | 0,028005 | 0,002647 |
| 4 | 6,126869 | 6,230071 | -0,103201 | -0,11066 | -0,91085 | 0,057845 | 1,950434 | -0,139583 | 0,098896 |
| 5 | 6,111467 | 6,145164 | -0,033696 | -0,40860 | -0,29740 | 0,064190 | 2,613880 | -0,049623 | 0,015391 |
| 6 | 6,398595 | 6,238333 | 0,160262 | -0,08167 | 1,41446 | 0,056269 | 1,796330 | 0,212729 | 0,217355 |
| 7 | 6,408529 | 6,458878 | -0,050349 | 0,69223 | -0,44438 | 0,084714 | 5,232613 | -0,114177 | 0,141923 |
| 8 | 6,371612 | 6,297143 | 0,074470 | 0,12470 | 0,65726 | 0,052807 | 1,472806 | 0,095135 | 0,038287 |
| 9 | 6,171700 | 6,342323 | -0,170623 | 0,28324 | -1,50591 | 0,049310 | 1,166792 | -0,210491 | 0,163426 |
| 10 | 6,529419 | 6,427354 | 0,102065 | 0,58161 | 0,90082 | 0,039615 | 0,428038 | 0,116279 | 0,032188 |
| 11 | 6,679599 | 6,636141 | 0,043458 | 1,31425 | 0,38356 | 0,068129 | 3,060506 | 0,068070 | 0,032625 |
| 12 | 6,664409 | 6,670452 | -0,006043 | 1,43465 | -0,05333 | 0,084165 | 5,153168 | -0,013483 | 0,001953 |
| Minimum | 5,713733 | 5,648688 | -0,170623 | -2,15075 | -1,50591 | 0,039615 | 0,428038 | -0,210491 | 0,001953 |
| Maximum | 6,679599 | 6,670452 | 0,160262 | 1,43465 | 1,41446 | 0,095807 | 6,948477 | 0,228238 | 0,725357 |
| Average | 6,261607 | 6,261606 | 0,000000 | 0,00000 | 0,00000 | 0,063350 | 2,750000 | 0,006290 | 0,132213 |
| Median | 6,271656 | 6,267737 | 0,008554 | 0,02152 | 0,07549 | 0,059014 | 2,068645 | 0,007261 | 0,068591 |

$$\sigma_y = 97 + 147C + 5{,}07Si - 0{,}4t + 0{,}033Q - 2{,}84\upsilon \quad (8)$$

$$\sigma_b = 123 + 111C + 1{,}56Si - 0{,}3t + 0{,}056Q - 2{,}98\upsilon \quad (9)$$

$$\delta_5 = 28{,}0 - 9{,}2C + 2{,}11Si - 0{,}018t - 0{,}004Q - 0{,}92\upsilon \quad (10)$$

where t - temperature of water cooling; Q - water discharge; $\upsilon$ - rolling rate.
The analysis of these equations shows, that the finding nature of dependence, defined sign of coefficients, meets physical sense one or another technological factor to mechanical properties of strain-heat hardened steel. So, the increase of carbon content in steel causes the growth of $\sigma_b$, $\sigma_y$ and drop of $\delta_5$. Silicon, as well as the carbon hardenly influence on steel, practically do not make worth of plasticity. The increase of water discharge raises rate of cooling of steels and its strength properties, which one are decreased with growth of temperature of cooling water. The increase of rolling rate reduces of a cooling time, and rises temperature of self-tempering, as a consequence, to drop of strength properties. It is possible to explain some exceptions, for example, influencing of rolling rate on $\sigma_b$, $\sigma_y$ and $\delta_5$, by correlating action of other factors.